\documentclass[pre,amsmath,amssymb,reprint]{revtex4-2}
\usepackage{epsfig}                                                                 
\usepackage{dcolumn}                                                                
\usepackage{bm}                                                                     
\usepackage{xcolor}                                                                 

\usepackage[pdfstartview=FitH,bookmarks=false]{hyperref}                            

\begin{document}
\title[Can a liquid drop on a substrate be in equilibrium with saturated vapor?]%
{Can a liquid drop on a substrate be in equilibrium with saturated vapor?}
\author{E. S. Benilov}
 \altaffiliation[]{Department of Mathematics and Statistics, University of Limerick, Limerick, V94 T9PX, Ireland}
 \email{Eugene.Benilov@ul.ie}
 \homepage{https://staff.ul.ie/eugenebenilov/}
\date{\today}

\begin{abstract}
It is well-known that liquid and saturated vapor, separated by a flat
interface in an unbounded space, are in equilibrium. One would similarly
expect a liquid drop, sitting on a flat substrate, to be in equilibrium with
the vapor surrounding it. Yet, it is not: as shown in this work, the drop
evaporates. Mathematically, this conclusion is deduced using the
diffuse-interface model, but it is also reformulated in terms of the
maximum-entropy principle, suggesting model independence. Physically,
evaporation of drops is due to the so-called Kelvin effect, which gives rise
to a liquid-to-vapor mass flux if the boundary of the liquid phase is convex.

\end{abstract}
\maketitle

\noindent\emph{Introduction. }The diffuse-interface model (DIM) was proposed
by Korteweg in 1901 \cite{Korteweg01}, and later developed by Ginzburg
\cite{Ginzburg60} and Cahn \cite{Cahn61}. Since then, it has been used in
numerous problems including nucleation and collapse of bubbles
\cite{MagalettiMarinoCasciola15,MagalettiGalloMarinoCasciola16,GalloMagalettiCasciola18,GalloMagalettiCoccoCasciola20}%
, phase separation in polymer blends
\cite{ThieleMadrugaFrastia07,MadrugaThiele09}, contact lines
\cite{DingSpelt07,YueZhouFeng10,YueFeng11,SibleyNoldSavvaKalliadasis13a,SibleyNoldSavvaKalliadasis13b,SibleyNoldSavvaKalliadasis14,BorciaBestehorn14,BorciaBorciaBestehornVarlamovaHoefnerReif19}%
, etc.

It was also used in Ref. \cite{Benilov20c} to prove the nonexistence of
solutions describing static two-dimensional (2D) drops on a substrate. This
result, however, gives rise to numerous follow-up questions. If a drop cannot
be static, how exactly does it evolve? Is it spreading out, while decreasing
in thickness -- or perhaps it acts as a center of condensation for the
surrounding vapor and, thus, grows? In addition, the 2D proof of Ref.
\cite{Benilov20c} was not applicable to the most interesting case, that of 3D
axisymmetric drops. Does this mean that they can be static -- or their
nonexistence can be still proved using a different method?

In the present work, the approach of Ref. \cite{Benilov20c} is modified to
prove the nonexistence of static 3D drops on a substrate. This result is then
reformulated in terms of the maximum entropy principle and interpreted via the
Kelvin effect
\cite{EggersPismen10,ColinetRednikov11,RednikovColinet13,Morris14,JanecekDoumencGuerrierNikolayev15,SaxtonWhiteleyVellaOliver16,RednikovColinet17,RednikovColinet19}%
, with both suggesting that drops \emph{evaporate}. This conclusion agrees
with, and explains, the evaporation of drops observed in numerical simulations
\cite{YueZhouFeng07}.\smallskip

\noindent\emph{Formulation. }There are two different versions of the
diffuse-interface model (DIM): one assuming the fluid velocity to be
solenoidal \cite{HohenbergHalperin77,JasnowVinals96} and another, based on the
full equations of compressible hydrodynamics
\cite{AndersonMcFaddenWheeler98,PismenPomeau00}. For static solutions,
however, the two models coincide.

Even though the results below are applicable to an arbitrary nonideal fluid,
they are easier to present using the van der Waals equation of state (say,
with parameters $a$ and $b$). Introduce also the Korteweg constant $K$
characterizing the intermolecular attractive force
\cite{Korteweg01,HohenbergHalperin77,JasnowVinals96,AndersonMcFaddenWheeler98,PismenPomeau00}%
.

The following nondimensional variables (marked with the subscript $_{nd}$)
will be used:%
\[
r_{nd}=\frac{r}{l},\qquad z_{nd}=\frac{z}{l},\qquad\rho_{nd}=b\rho,\qquad
T_{nd}=\frac{RTb}{a},
\]
where $r$ is the horizontal (polar) radius, $z$ is the vertical (axial)
coordinate, $\rho$ is the density, $T$ is the temperature, $R$ is the specific
gas constant, and $l=\left(  K/a\right)  ^{1/2}$ is the characteristic
interfacial thickness. Physically, $l$ is on a nanometer scale and, thus, will
be referred to as \textquotedblleft microscopic\textquotedblright.

According to the DIM, a static axisymmetric distribution of a van der Waals
fluid satisfies (the subscript $_{nd}$ omitted)%
\begin{multline}
\frac{1}{r}\frac{\partial}{\partial r}\left(  r\frac{\partial\rho}{\partial
r}\right)  +\frac{\partial^{2}\rho}{\partial z^{2}}\\
=T\left(  \ln\frac{\rho}{1-\rho}+\frac{1}{1-\rho}\right)  -2\rho-\mu,
\label{1}%
\end{multline}
where $\mu$ is a constant. The physical meaning of this equation will be
explained later.

\begin{figure}
\includegraphics[width=\columnwidth]{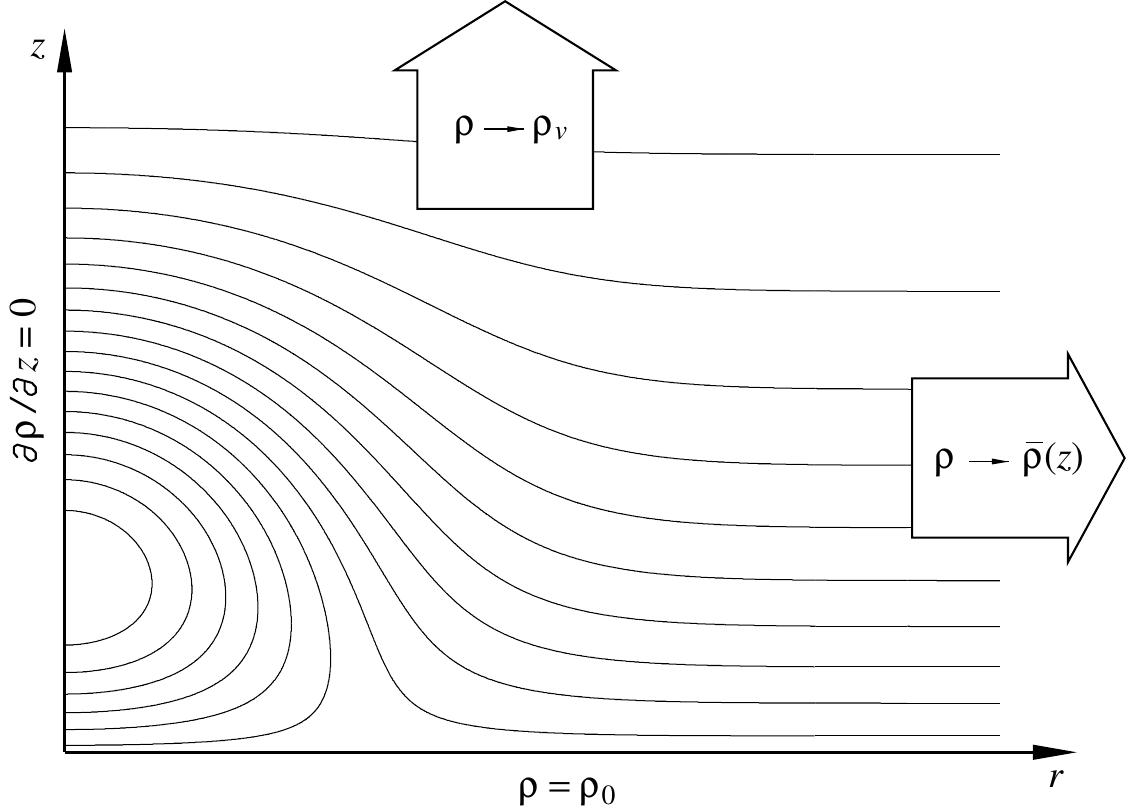}
\caption{A schematic illustrating the setup of boundary conditions for, and a typical solution of, Eq. (\ref{1}).}\label{fig1}
\end{figure}

As illustrated in Fig. \ref{fig1}, Eq. (\ref{1}) requires four boundary conditions.

Let the drop be horizontally and vertically localized. The latter implies
that, far above the substrate, the density tends to a constant -- say,
$\rho_{\infty}$. If $\rho_{\infty}$ exceeds the density $\rho_{v}$ of
saturated vapor, the setting under consideration becomes physically
meaningless (because an infinitely large volume of \emph{over}saturated vapor
-- with or without a liquid drop -- is thermodynamically unstable). If, in
turn, $\rho_{\infty}<\rho_{v}$, the problem becomes trivial, as evaporation of
drops surrounded by \emph{under}saturated vapor is evident without proof.
Thus, assume%
\begin{equation}
\rho\rightarrow\rho_{v}\qquad\text{as}\qquad z\rightarrow\infty.\label{2}%
\end{equation}
The saturated-vapor density $\rho_{v}$ can only be defined together with the
matching liquid density $\rho_{l}$ (in what follows, the latter will not be
involved). They are both determined by the Maxwell construction, comprising
the requirements that the vapor's pressure and chemical potential match those
of the liquid. For the van der Waals fluid, these requirements amount to%
\begin{equation}
\frac{T\rho_{v}}{1-\rho_{v}}-\rho_{v}^{2}=\frac{T\rho_{l}}{1-\rho_{l}}%
-\rho_{l}^{2},\label{3}%
\end{equation}%
\begin{multline}
T\left(  \ln\frac{\rho_{v}}{1-\rho_{v}}+\frac{1}{1-\rho_{v}}\right)
-2\rho_{v}\\
=T\left(  \ln\frac{\rho_{l}}{1-\rho_{l}}+\frac{1}{1-\rho_{l}}\right)
-2\rho_{l}.\label{4}%
\end{multline}
It can be shown that, if $T<8/27$ (subcritical temperature for the van der
Waals fluid), Eqs. (\ref{3})--(\ref{4}) admit a unique solution such that
$\rho_{l}>\rho_{v}$ and%
\begin{equation}
\frac{T}{\left(  1-\rho_{v}\right)  ^{2}\rho_{v}}>2,\qquad\frac{T}{\left(
1-\rho_{l}\right)  ^{2}\rho_{l}}>2.\label{5}%
\end{equation}
Conditions (\ref{5}) guarantee that both liquid and vapor are stable (the
pressure grows with density). For supercritical temperatures, only the trivial
solution ($\rho_{l}=\rho_{v}$) exists, so interfaces do not.

Observe that boundary condition (\ref{2}) and Eq. (\ref{1}) entail%
\begin{equation}
\mu=T\left(  \ln\frac{\rho_{v}}{1-\rho_{v}}+\frac{1}{1-\rho_{v}}\right)
-2\rho_{v}.\label{6}%
\end{equation}
Physically, $\mu$ is the specific chemical potential of the van der Waals
vapor (or, to be precise, differs from that by a constant).

Let the fluid be bounded below by a substrate located at $z=0$, in which case
the DIM implies \cite{PismenPomeau00,Benilov20a}%
\begin{equation}
\rho\rightarrow\rho_{0}\qquad\text{as}\qquad z\rightarrow0,\label{7}%
\end{equation}
where $\rho_{0}$ is a constant characterizing the liquid/substrate
interaction. In this paper, the substrate is assumed to be neither perfectly
hydrophilic nor perfectly hydrophobic, with the implication that $\rho
_{v}<\rho_{0}<\rho_{l}$ \cite{PismenPomeau00}.

To clarify the physical meaning of condition (\ref{7}), consider the
intermolecular forces exerted on a fluid molecule in an infinitesimally-thin
layer near the substrate: the solid pulls the molecule toward the substrate,
while the fluid outside the layer pulls it in the opposite direction. Since
the former force is fixed, whereas the latter grows monotonically with the
near-substrate density, the balance is achieved when the density assumes a
certain value -- which is what condition (\ref{7}) reflects. Furthermore, the
main conclusion of this paper (evaporation of all sessile drops) would not
change even if (\ref{7}) were replaced by the Neumann or mixed boundary
conditions (assumed phenomenologically and used, for example, in Refs.
\cite{Cahn61,Seppecher96} and \cite{ThieleMadrugaFrastia07,MadrugaThiele09},
respectively). From a physical viewpoint, evaporation occurs at the
liquid/vapor interface, so the fluid--substrate interaction does not affect it much.

The fact that the drop is localized horizontally implies that, at large $r$,
the substrate is dry -- i.e., the vapor and solid are adjacent to one another,
with no liquid in between. In terms of the DIM, a solid/vapor interface
corresponds to a microscopic boundary layer where the density field is
homogeneous horizontally, but changes vertically from $\rho_{0}$ at the
substrate toward $\rho_{v}$ far above it. The profile $\bar{\rho}(z)$ of this
layer satisfies the one-dimensional reduction of boundary-value problem
(\ref{1}), (\ref{7}), (\ref{2}),%
\begin{equation}
\frac{\mathrm{d}^{2}\bar{\rho}}{\mathrm{d}z^{2}}=T\left(  \ln\frac{\bar{\rho}%
}{1-\bar{\rho}}+\frac{1}{1-\bar{\rho}}\right)  -2\bar{\rho}-\mu, \label{8}%
\end{equation}%
\begin{equation}
\bar{\rho}=\rho_{0}\qquad\text{at}\qquad z=0, \label{9}%
\end{equation}%
\begin{equation}
\bar{\rho}\rightarrow\rho_{v}\qquad\text{as}\qquad z\rightarrow\infty.
\label{10}%
\end{equation}
Thus, horizontal localization of the drop assumes that%
\begin{equation}
\rho\rightarrow\bar{\rho}(z)\qquad\text{as}\qquad r\rightarrow\infty.
\label{11}%
\end{equation}
To ensure that $\rho(r,z)$ is analytic at $r=0$, require%
\begin{equation}
\frac{\partial\rho}{\partial r}=0\qquad\text{at}\qquad r=0. \label{12}%
\end{equation}
Finally, let%
\begin{equation}
\left\vert \int_{0}^{\infty}\left(  \rho-\bar{\rho}\right)  r\,\mathrm{d}%
r\right\vert <\infty, \label{13}%
\end{equation}
which implies that the excess mass between any two horizontal planes is
finite, and so is the drop's net mass.\smallskip

\noindent\emph{Properties of boundary-value problem (\ref{1})--(\ref{13}).
}Mathematically, Eqs. (\ref{1})--(\ref{13}) have a lot in common with their 2D
counterparts examined in Ref. \cite{Benilov20c}. In what follows, the
properties of the former will be briefly outlined with references to the latter.

(i) As $z\rightarrow\infty$, Eq. (\ref{1}) can be linearized against the
background of $\rho_{v}$ and thus becomes a Helmholtz equation,
\begin{multline}
\frac{1}{r}\frac{\partial}{\partial r}\left[  r\frac{\partial\left(  \rho
-\rho_{v}\right)  }{\partial r}\right]  +\frac{\partial^{2}\left(  \rho
-\rho_{v}\right)  }{\partial z^{2}}\\
-\left[  \frac{T}{\left(  1-\rho_{v}\right)  ^{2}\rho_{v}}-2\right]  \left(
\rho-\rho_{v}\right)  =0.\label{14}%
\end{multline}
By virtue of (\ref{5}), the second expression in the square brackets is
positive -- which implies that all solutions of Eq. (\ref{14}) are either
exponentially decaying or exponentially growing as $z\rightarrow\infty$. The
latter is ruled out by boundary condition (\ref{2}) -- hence,%
\begin{equation}
\left(  \rho-\rho_{v}\right)  z^{n}\rightarrow0,\hspace{0.5cm}\frac
{\partial\rho}{\partial z}z^{n}\rightarrow0\hspace{0.5cm}\text{as}%
\hspace{0.5cm}z\rightarrow\infty,\label{15}%
\end{equation}
for all $n$.

The above argument can be reworked into a formal proof similar to that for 2D
drops in Ref. \cite{Benilov20c}. One only needs to replace in the latter the
Fourier transformation with the Hankel transformation.

(ii) As $r\rightarrow\infty$, Eq. (\ref{1}) can be linearized against the
background of $\bar{\rho}(z)$ and written in the form%
\begin{equation}
\frac{1}{r}\frac{\partial}{\partial r}\left[  r\frac{\partial\left(  \rho
-\bar{\rho}\right)  }{\partial r}\right]  -\hat{A}\left(  \rho-\bar{\rho
}\right)  =0,\label{16}%
\end{equation}
where the operator%
\begin{equation}
\hat{A}=-\frac{\partial^{2}}{\partial z^{2}}+\left[  \frac{T}{\left(
1-\bar{\rho}\right)  ^{2}\bar{\rho}}-2\right]  \label{17}%
\end{equation}
is self-adjoint. As before, (\ref{16})--(\ref{17}) form a Helmholtz equation,
but this time the expression in the square brackets in (\ref{17}) can be
negative for some $z$. As a result, the exponential decay of the solutions of
Eq. (\ref{16}) as $r\rightarrow$ $\infty$ is not obvious, but still follows
from the fact that the operator $\hat{A}$ is positive-definite (see Lemma 4 of
Ref. \cite{Benilov20c}). Thus,%
\begin{equation}
\left(  \rho-\bar{\rho}\right)  r^{n}\rightarrow0,\hspace{0.5cm}\frac
{\partial\rho}{\partial r}r^{n}\rightarrow0\hspace{0.5cm}\text{as}%
\hspace{0.5cm}r\rightarrow\infty\label{18}%
\end{equation}
for all $n$.

(iii) Assume that $\varrho(z)$ is a smooth function such that $\varrho
(0)=\rho_{0}$ and $\varrho(z)\rightarrow\rho_{v}$ as $z\rightarrow\infty$, and
introduce the following functional%
\begin{multline*}
F[\varrho(z)]=\int_{0}^{\infty}\left[  \frac{1}{2}\left(  \frac{\mathrm{d}%
\varrho}{\mathrm{d}z}\right)  ^{2}\right. \\
+\left.
\vphantom{\left( \frac{\mathrm{d}\varrho}{\mathrm{d}\varrho }\right) ^{2}}T\varrho
\ln\frac{\varrho}{1-\varrho}-\varrho^{2}-\mu\varrho+p\right]  \mathrm{d}z
\end{multline*}
where $\mu$ is given by (\ref{6}) and%
\[
p=\frac{T\rho_{v}}{1-\rho_{v}}-\rho_{v}^{2}%
\]
is, physically, the pressure of saturated van-der-Waals vapor.

As shown in Ref. \cite{Benilov20c}, $F$ reaches the absolute minimum when
$\varrho=\bar{\rho}(z)$. Thus, if (\ref{1})--(\ref{13}) admit a non-trivial
($\rho\neq\bar{\rho}$) solution, it satisfies%
\begin{equation}
F[\rho(r,z)]>F[\bar{\rho}(z)]\qquad\forall r\in\left(  0,\infty\right)  .
\label{19}%
\end{equation}
Note that the integrals involved in $F[\bar{\rho}(z)]$ converge due to
(\ref{15}).\smallskip

\noindent\emph{Nonexistence of static drops. }The nonexistence of solutions of
boundary-value problem (\ref{1})--(\ref{13}) will be proved by contradiction.

Assuming that a solution exists, multiply Eq. (\ref{1}) by $r^{2}\partial
\rho/\partial r$ and integrate from $z=0$ to $z=\infty$. Integrating by parts
and using conditions (\ref{15}) to interchange differentiation with respect to
$r$ and integration with respect to $z$, one obtains%
\[
r^{2}\frac{\partial F[\rho(r,z)]}{\partial r}-\frac{1}{2}\frac{\partial
}{\partial r}\int_{0}^{\infty}\left(  r\frac{\partial\rho}{\partial r}\right)
^{2}\mathrm{d}z=0.
\]
Integrate this equality with respect to $r$ from $r=0$ to $r=\infty$ and,
replacing in the first term%
\[
\frac{\partial F[\rho(r,z)]}{\partial r}\rightarrow\frac{\partial\left(
F[\rho(r,z)]-F[\bar{\rho}(z)]\right)  }{\partial r},
\]
integrate by parts. Recalling conditions (\ref{18}), one obtains%
\[
\int_{0}^{\infty}r\left\{  F[\rho(r,z)]-F[\bar{\rho}(z)]\right\}
\mathrm{d}r=0.
\]
Given (\ref{19}), this last equality is incorrect -- hence, the contradiction.

Thus, the only existing steady-state solution is the trivial one (describing
dry substrate).\smallskip

\noindent\emph{The maximum entropy principle.} The steady-state equations
(\ref{1})--(\ref{13}) can be reformulated as a problem of maximization of the
net entropy subject to the net energy and mass being fixed.

To this end, introduce the specific (per unit mass) entropy $s(\rho,T)$ and
the specific internal energy $e(\rho,T)$. They are not entirely arbitrary, as
they are supposed to satisfy the fundamental thermodynamic relation,%
\begin{equation}
\frac{\partial e}{\partial T}=T\frac{\partial s}{\partial T}.\label{20}%
\end{equation}
Introduce also the fluid's chemical potential,%
\begin{equation}
G=\frac{\partial}{\partial\rho}\left[  \rho\left(  e-Ts\right)  \right]
.\label{21}%
\end{equation}
For the van der Waals fluid, for example, one has (nondimensionally)%
\[
e=c_{V}T-\rho,\qquad s=c_{V}\ln T-\ln\frac{\rho}{1-\rho},
\]%
\[
G=T\left(  \ln\frac{\rho}{1-\rho}+\frac{1}{1-\rho}\right)  -2\rho+c_{V}\left(
1-\ln T\right)  ,
\]
where $c_{V}$ is the nondimensional heat capacity at constant volume.

The van der Waals equation (\ref{1}) can now be extended to the general case,
in the form%
\begin{equation}
\nabla^{2}\rho=G-\mu^{\prime}, \label{22}%
\end{equation}
where $\nabla^{2}$ is the axisymmetric Laplace operator and $\mu^{\prime}$ is
related to its van der Waals counterpart by $\mu^{\prime}=\mu+c_{V}\left(
1-\ln T\right)  $.

Next, introduce the net excess mass $M$, the net excess entropy $S$, and the
net excess full energy $E$,%
\[
M=\int\left(  \rho-\bar{\rho}\right)  \mathrm{d}V,\qquad S=\int\left(  \rho
s-\bar{\rho}\bar{s}\right)  \,\mathrm{d}V,
\]%
\begin{equation}
E=\int\left\{  \rho e-\bar{\rho}\bar{e}+\frac{1}{2}\left[  \left(
\frac{\partial\rho}{\partial r}\right)  ^{2}+\left(  \frac{\partial\rho
}{\partial z}\right)  ^{2}\right]  \right\}  \mathrm{d}V, \label{23}%
\end{equation}
where $\bar{s}=s(\bar{\rho},T)$, $\bar{e}=e(\bar{\rho},T)$, $\mathrm{d}V=2\pi
r\,\mathrm{d}r\,\mathrm{d}z$, and the integrals are evaluated over the
semispace $z>0$. The derivative terms in expression (\ref{23}) represent the
energy of the intermolecular attraction as the DIM describes it.

All of the results obtained earlier for the van der Waals equation (\ref{1})
can be readily reproduced for the general equation (\ref{22}). Even more
importantly, the latter can be used to find out the physical meaning of the
nonexistence of solutions describing sessile drops.

Observe that Eq. (\ref{22}) follows from the requirement that the net entropy
be maximum subject to the net mass and energy be fixed, i.e.,%
\[
\delta\left(  S+\eta M+\lambda E\right)  =0,
\]
where $\eta$ and $\lambda$ are the Langrange multipliers. Carrying out
variation in the above equation and recalling boundary conditions (\ref{2})
and (\ref{7}), which imply%
\[
\delta\rho=0\qquad\text{at}\qquad z=0,
\]%
\[
\delta\rho\rightarrow0\qquad\text{as}\qquad z\rightarrow\infty,
\]
one obtains%
\begin{multline*}
\int\left[  \frac{\partial\left(  \rho s\right)  }{\partial\rho}+\lambda
\frac{\partial\left(  \rho e\right)  }{\partial\rho}-\lambda\nabla^{2}%
\rho+\eta\right]  \delta\rho\,\mathrm{d}V\\
+\int\left[  \frac{\partial\left(  \rho s\right)  }{\partial T}+\lambda
\frac{\partial\left(  \rho e\right)  }{\partial T}\right]  \delta
T\,\mathrm{d}V=0.
\end{multline*}
Setting $\lambda=-1/T$, one can make the second term in this equation vanish
subject to identity (\ref{20}), whereas $\eta=\mu^{\prime}/T$ makes the first
term vanish subject to condition (\ref{21}) and the steady-state equation
(\ref{22}).

Thus, since $\bar{\rho}(z)$ (describing the solid/vapor interface) is the only
solution of Eq. (\ref{22}), it corresponds to the maximum entropy and all
other solutions evolve towards it. This means evaporation of the liquid phase,
with the excess mass and energy spreading out to infinity.

The only DIM-specific part of the above variational problem is the
intermolecular part of energy (\ref{23}). It seems unlikely that another form
of this term would fundamentally change the properties of the functionals
involved. Hence, one could conjecture that drops on a solid substrate
evaporate in \emph{any} model conserving mass and energy, and conforming to an
H-Theorem -- such as, for example, the Enskog--Vlasov kinetic equation for
dense fluids
\cite{Desobrino67,Grmela71,GrmelaGarciacolin80,BarbanteFrezzottiGibelli15,BenilovBenilov18,BenilovBenilov19b}
(which is generally viewed as a much more accurate model than the
DIM).\smallskip

\noindent\emph{Physical interpretation.} The nonexistence of drops on a
substrate can be explained through the so-called Kelvin effect
\cite{EggersPismen10,ColinetRednikov11,RednikovColinet13,Morris14,JanecekDoumencGuerrierNikolayev15,RednikovColinet17,RednikovColinet19}%
, which gives rise to a mass flux through a liquid/vapor interface provided it
is curved. The direction of the flux depends on the sign of the interfacial
curvature: for a volume of liquid with a \emph{convex} boundary, the flux is
directed \emph{from} the liquid \emph{toward} the vapor, and vice versa. Since
the boundary of a drop on a flat substrate is convex, it comes as no surprise
that it evaporates. One can further conjecture that drops floating in
saturated vapor evaporate too \cite{Benilov21d}. As shown in Ref.
\cite{Benilov21c}, the only kind of drops that do not evaporate are those in a
sufficiently acute corner, so that their surface is concave -- hence, vapor is
condensating on it.

Finally, the quantitative theory of the Kelvin effect can be used to show
that, depending on the drop size and temperature, the timescale of evaporation
ranges from several seconds to several days. This estimate (to be published
separately) suggests that the evaporation of drops into saturated vapor can be
observed experimentally, and it should certainly be observable numerically via
molecular dynamics.

\bibliography{}

\end{document}